\documentclass[11pt]{article}
\usepackage{graphicx}
\usepackage{amssymb,amsmath}
\usepackage{xcolor}
\usepackage[sort&compress,numbers]{natbib}

\newcommand{\tadpole}{
\begin{picture}(16,14)
\put(8,0){\circle*{3}}
\put(8,6){\circle{12}}
\end{picture}  }
\newcommand{\tadpolex}{
\begin{picture}(16,14)
\put(8,0){\circle*{3}}
\put(8,6){\circle{12}}
\bezier{40}(2,6)(5,9)(8,12)
\bezier{40}(8,0)(11,3)(14,6)
\bezier{60}(2.8,3.8)(5,6)(10.2,11.2)
\bezier{60}(5.8,0.8)(11,6)(13.2,8.2)
\bezier{60}(3.8,1.8)(8,6)(12.2,10.2)
\end{picture}  }
\newcommand{\tadpoley}{
\begin{picture}(18,14)
\put(9,0){\circle*{3}}
\put(3,6){\circle*{3}}
\put(15,6){\circle*{3}}
\put(9,6){\circle{12}}
\bezier{60}(3,6)(9,10)(15,6)
\bezier{60}(3,6)(9,2)(15,6)
\end{picture}  }
\newcommand{\selfekk}{
\begin{picture}(44,14)
\thicklines
\put(21,3.5){\circle{11}}
\multiput(20.8,-1.6)(0.5,0.5){12}{\circle*{1}}
\multiput(21.2,8.6)(-0.5,-0.5){12}{\circle*{1}}
\multiput(18.2,-1)(0.5,0.5){15}{\circle*{1}}
\multiput(23.8,8)(-0.5,-0.5){15}{\circle*{1}}
\put(15.5,3.5){\line(-1,0){12}}
\put(26.5,3.5){\line(1,0){14}}
\put(3.5,5.5){\footnotesize {\bf k}}
\put(32,5.5){\footnotesize -{\bf k}}
\end{picture} }
\newcommand{\selfeko}{
\begin{picture}(33,14)
\thicklines
\put(3.5,3.5){\line(1,0){26}}
\put(3.5,5.5){\footnotesize {\bf k}}
\put(21,5.5){\footnotesize -{\bf k}}
\end{picture} }
\newcommand{\selfekkk}{
\begin{picture}(65,14)
\thicklines
\multiput(21,3.5)(21,0){2}{\circle{11}}
\multiput(20.8,-1.6)(0.5,0.5){12}{\circle*{1}}
\multiput(21.2,8.6)(-0.5,-0.5){12}{\circle*{1}}
\multiput(18.2,-1)(0.5,0.5){15}{\circle*{1}}
\multiput(23.8,8)(-0.5,-0.5){15}{\circle*{1}}
\put(26.5,3.5){\line(1,0){10}}
\multiput(41.8,-1.6)(0.5,0.5){12}{\circle*{1}}
\multiput(42.2,8.6)(-0.5,-0.5){12}{\circle*{1}}
\multiput(39.2,-1)(0.5,0.5){15}{\circle*{1}}
\multiput(44.8,8)(-0.5,-0.5){15}{\circle*{1}}
\put(15.5,3.5){\line(-1,0){12}}
\put(47.5,3.5){\line(1,0){14}}
\put(3.5,5.5){\footnotesize {\bf k}}
\put(53,5.5){\footnotesize -{\bf k}}
\end{picture} }
\newcommand{\wl}{
\begin{picture}(50,12)
\multiput(7,0)(8,0){5}{\oval(4,4)[t]}
\multiput(11,0)(8,0){5}{\oval(4,4)[b]}
\put(5,3.5){\footnotesize {\bf k}}
\put(37,3.5){\footnotesize -{\bf k}}
\end{picture}  }
\newcommand{\Dthree}{
\begin{picture}(47,12)
\put(16,4){\line(-1,0){14}}
\put(26,4){\oval(20,12)}
\put(28,10){\line(0,-1){12}}
\put(36,4){\circle*{3}}
\put(36,4){\line(-1,0){8}}
\put(38,0){\footnotesize -{\bf k}}
\bezier{20}(6,6)(8,4)(10,2)
\bezier{20}(6,2)(8,4)(10,6)
\bezier{40}(16,4)(19,7)(22,10)
\bezier{60}(16.5,1.5)(22,7)(25,10)
\bezier{60}(18,0)(22,4)(28,10)
\bezier{60}(19.7,-1.3)(22,1)(28,7)
\bezier{40}(22.2,-1.8)(25,1)(28,4)
\bezier{20}(25,-2)(26.5,-0.5)(28,1)
\end{picture}
}
\newcommand{\Dtthree}{
\begin{picture}(50,12)
\put(25,4){\oval(28,12)}
\multiput(11,4)(28,0){2}{\circle*{3}}
\put(3,0){\footnotesize {\bf k}}
\put(41,0){\footnotesize -{\bf k}}
\multiput(19,10)(12,0){2}{\line(0,-1){12}}
\multiput(11,4)(20,0){2}{\line(1,0){8}}
\bezier{20}(19,6)(21,8)(23,10)
\bezier{40}(19,2)(23,6)(27,10)
\bezier{60}(19,-2)(25,4)(31,10)
\bezier{40}(23,-2)(27,2)(31,6)
\bezier{20}(27,-2)(29,0)(31,2)
\end{picture}
}
\newcommand{\wltad}{
\begin{picture}(52,17)
\multiput(16,2)(8,0){3}{\oval(4,4)[t]}
\multiput(20,2)(8,0){3}{\oval(4,4)[b]}
\put(5.5,0){\footnotesize {\bf k}}
\put(14,2){\circle*{3}}
\put(39,3){\circle*{3}}
\put(39,9){\circle{12}}
\bezier{40}(33,9)(36,12)(39,15)
\bezier{40}(39,3)(42,6)(45,9)
\bezier{60}(33.8,6.8)(36,9)(41.2,14.2)
\bezier{60}(36.8,3.8)(42,9)(44.2,11.2)
\bezier{60}(34.8,4.8)(39,9)(43.2,13.2)
\put(43,0){\footnotesize -{\bf k}}
\end{picture}  }
\newcommand{\wlDthree}{
\begin{picture}(69,15)
\multiput(14,4)(8,0){3}{\oval(4,4)[t]}
\multiput(18,4)(8,0){3}{\oval(4,4)[b]}
\put(3.5,0){\footnotesize {\bf k}}
\put(12,4){\circle*{3}}
\put(46,4){\oval(20,12)}
\put(48,10){\line(0,-1){12}}
\put(56,4){\circle*{3}}
\put(56,4){\line(-1,0){8}}
\put(58.5,0){\footnotesize -{\bf k}}
\bezier{40}(36,4)(39,7)(42,10)
\bezier{60}(36.5,1.5)(42,7)(45,10)
\bezier{60}(38,0)(42,4)(48,10)
\bezier{60}(39.7,-1.3)(42,1)(48,7)
\bezier{40}(42.2,-1.8)(45,1)(48,4)
\bezier{20}(45,-2)(46.5,-0.5)(48,1)
\end{picture}  }
\newcommand{\twotad}{
\begin{picture}(52,17)
\multiput(24,2)(8,0){2}{\oval(4,4)[t]}
\multiput(20,2)(8,0){3}{\oval(4,4)[b]}
\put(6,0){\footnotesize {\bf k}}
\put(17,3){\circle*{3}}
\put(39,3){\circle*{3}}
\put(39,9){\circle{12}}
\bezier{40}(33,9)(36,12)(39,15)
\bezier{40}(39,3)(42,6)(45,9)
\bezier{60}(33.8,6.8)(36,9)(41.2,14.2)
\bezier{60}(36.8,3.8)(42,9)(44.2,11.2)
\bezier{60}(34.8,4.8)(39,9)(43.2,13.2)
\put(17,9){\circle{12}}
\bezier{40}(11,9)(14,12)(17,15)
\bezier{40}(17,3)(20,6)(23,9)
\bezier{60}(11.8,6.8)(14,9)(19.2,14.2)
\bezier{60}(14.8,3.8)(20,9)(22.2,11.2)
\bezier{60}(12.8,4.8)(17,9)(21.2,13.2)
\put(43,0){\footnotesize -{\bf k}}
\end{picture}  }
\newcommand{\tadDthree}{
\begin{picture}(70,25)
\multiput(23,4)(8,0){2}{\oval(4,4)[t]}
\multiput(19,4)(8,0){3}{\oval(4,4)[b]}
\put(6,0){\footnotesize {\bf k}}
\put(16,5){\circle*{3}}
\put(47,4){\oval(20,12)}
\put(49,10){\line(0,-1){12}}
\put(57,4){\circle*{3}}
\put(57,4){\line(-1,0){8}}
\put(59.5,0){\footnotesize -{\bf k}}
\bezier{40}(37,4)(40,7)(43,10)
\bezier{60}(37.5,1.5)(43,7)(46,10)
\bezier{60}(39,0)(43,4)(49,10)
\bezier{60}(40.7,-1.3)(43,1)(49,7)
\bezier{40}(43.2,-1.8)(46,1)(49,4)
\bezier{20}(46,-2)(47.5,-0.5)(49,1)
\put(16,11){\circle{12}}
\bezier{40}(10,11)(13,14)(16,17)
\bezier{40}(16,5)(19,8)(22,11)
\bezier{60}(10.8,8.8)(13,11)(18.2,16.2)
\bezier{60}(13.8,5.8)(19,11)(21.2,13.2)
\bezier{60}(11.8,6.8)(16,11)(20.2,15.2)
\end{picture}  }
\newcommand{\DwlD}{
\begin{picture}(81,12)
\multiput(34,4)(8,0){2}{\oval(4,4)[t]}
\multiput(38,4)(8,0){2}{\oval(4,4)[b]}
\put(3.5,0){\footnotesize {\bf k}}
\put(58,4){\oval(20,12)}
\put(60,10){\line(0,-1){12}}
\put(68,4){\circle*{3}}
\put(68,4){\line(-1,0){8}}
\put(70.5,0){\footnotesize -{\bf k}}
\bezier{40}(48,4)(51,7)(54,10)
\bezier{60}(48.5,1.5)(54,7)(57,10)
\bezier{60}(50,0)(54,4)(60,10)
\bezier{60}(51.7,-1.3)(54,1)(60,7)
\bezier{40}(54.2,-1.8)(57,1)(60,4)
\bezier{20}(57,-2)(58.5,-0.5)(60,1)
\put(22,4){\oval(20,12)}
\put(12,4){\circle*{3}}
\put(20,10){\line(0,-1){12}}
\put(12,4){\line(1,0){8}}
\bezier{40}(32,4)(29,1)(26,-2)
\bezier{60}(31.5,6.5)(26,1)(23,-2)
\bezier{60}(30,8)(26,4)(20,-2)
\bezier{60}(28.3,9.3)(26,7)(20,1)
\bezier{40}(25.8,9.8)(23,7)(20,4)
\bezier{20}(23,10)(21.5,8.5)(20,7)
\end{picture}  }
\newcommand{\Dfive}{
\begin{picture}(51,18)
\put(2,4){\line(1,0){38}}
\put(28,4){\circle{24}}
\put(39.5,4){\circle*{3}}
\put(42,0){\footnotesize -{\bf k}}
\bezier{20}(6,6)(8,4)(10,2)
\bezier{20}(6,2)(8,4)(10,6)
\put(33,4){\circle{12}}
\multiput(16,4)(11,0){2}{\circle*{3}}
\end{picture}
}
\newcommand{\Dfivex}{
\begin{picture}(51,18)
\put(2,4){\line(1,0){38}}
\put(28,4){\circle{24}}
\put(39.5,4){\circle*{3}}
\put(42,0){\footnotesize -{\bf k}}
\bezier{20}(6,6)(8,4)(10,2)
\bezier{20}(6,2)(8,4)(10,6)
\put(33,4){\circle{12}}
\multiput(16,4)(11,0){2}{\circle*{3}}
\put(30,-3.5){\line(1,3){5}}
\end{picture}
}
\newcommand{\sunsetz}{
\begin{picture}(27,12)
\put(4,0){\footnotesize {\bf 0}}
\put(18,4){\circle{12}}
\put(12,4){\line(1,0){12}}
\multiput(12,4)(12,0){2}{\circle*{3}}
\end{picture}
}
\newcommand{\sunsetk}{
\begin{picture}(27,12)
\put(4,0){\footnotesize {\bf k}}
\put(18,4){\circle{12}}
\put(12,4){\line(1,0){12}}
\multiput(12,4)(12,0){2}{\circle*{3}}
\end{picture}
}
\newcommand{\sunsetkx}{
\begin{picture}(17,12)
\put(8,4){\circle{12}}
\put(2,4){\line(1,0){12}}
\multiput(2,4)(12,0){2}{\circle*{3}}
\put(6,-3.5){\line(1,3){5}}
\end{picture}
}
\newcommand{\sixx}{
\begin{picture}(45,12)
\put(15,4){\line(-5,3){10}}
\put(15,4){\line(-5,-3){10}}
\put(15,4){\line(-1,0){10}}
\multiput(14,4)(17,0){2}{\circle*{2}}
\multiput(17.5,4)(2.5,0){5}{\circle*{1}}
\put(30,4){\line(5,3){10}}
\put(30,4){\line(5,-3){10}}
\put(30,4){\line(1,0){10}}
\end{picture}  }

\begin{document}

\title{
\textbf{Critical two--point correlation functions and  
``equation of motion'' in the $\varphi^4$ model}
}

\author{J. Kaupu\v{z}s
\thanks{E--mail: \texttt{kaupuzs@latnet.lv}} \hspace{1ex} \\ 
Institute of Mathematical Sciences and Information Technologies, \\
University of Liepaja, 14 Liela Street, Liepaja LV--3401, Latvia} 

\maketitle

\begin{abstract}
Critical two--point correlation functions in the continuous and lattice $\varphi^4$ models 
with scalar order parameter $\varphi$ are considered. We show by 
different non--perturbative methods that the critical correlation functions 
$\langle \varphi^n({\bf 0}) \varphi^m({\bf x}) \rangle$ are proportional to
$\langle \varphi({\bf 0}) \varphi({\bf x}) \rangle$ at $\mid {\bf x} \mid =x \to \infty$
for any positive odd integers $n$ and $m$. 
We investigate how our results and some other results for well--defined 
models can be 
related to the conformal field theory (CFT), considered by Rychkov and Tan, and reveal some problems here.
We find this CFT to be rather formal, as it is based on an ill--defined model.
Moreover, we find it very unlikely that the used there ``equation of motion'' 
really holds from the point of view of statistical physics.
\end{abstract}


\section{Introduction}

In this paper we investigate non--perturbatively the relations between different two--point correlation
functions in the scalar $\varphi^4$ model from the point of view of statistical physics. We consider
correlation functions $\langle \varphi^n({\bf 0}) \varphi^m({\bf x}) \rangle$ with positive integer
powers $n$ and $m$ of the local order parameter $\varphi({\bf x})$ at the critical point. For odd values of $n$ and $m$, we find that all
these correlation functions are asymptotically proportional to
$\langle \varphi({\bf 0}) \varphi({\bf x}) \rangle$ at large distances $\mid {\bf x} \mid \to \infty$.
Although this result may seem to be very simple, it has some fundamental importance when we try to
relate the results of the conformal field theory~\cite{RT2015,Showk,xx}
with those, which can be obtained purely from statistical physics.

We consider the continuous $\varphi^4$ model in the 
thermodynamic limit of diverging volume $V \to \infty$ with the Hamiltonian $\cal{H}$ given by
\begin{equation} \label{eq:H}
\frac{\cal{H}}{k_B T}= \int \left( r_0 \varphi^2({\bf x}) + c (\nabla \varphi({\bf x}))^2 
+ u \varphi^4({\bf x}) \right) d {\bf x} \;,
\end{equation}
where $\varphi({\bf x})$ is the scalar order parameter, depending on the
coordinate ${\bf x}$, $T$ is the temperature, $k_B$ is the Boltzmann constant,
whereas $r_0$, $c$ and $u$ are Hamiltonian parameters, which are functions of $T$.
It is assumed that there exists the upper cut-off parameter $\Lambda$ 
for the Fourier components of the order-parameter field $\varphi({\bf x})$. 
Namely, the Fourier--transformed Hamiltonian reads 
\begin{equation} \label{eq:Hf}
\frac{\cal{H}}{k_B T}= \sum\limits_{\bf k} \left( r_0+c \,{\bf k}^2 \right)
{\mid \varphi_{\bf k} \mid}^2 + uV^{-1}
\sum\limits_{{\bf k}_1,{\bf k}_2,{\bf k}_3}
\varphi_{{\bf k}_1} \varphi_{{\bf k}_2} \varphi_{{\bf k}_3}
\varphi_{-{\bf k}_1-{\bf k}_2-{\bf k}_3} \;,
\end{equation}
where $\varphi_{\bf k}=V^{-1/2} \int \varphi({\bf x}) \exp(-i {\bf kx}) \, d{\bf x}$
and $\varphi({\bf x}) = V^{-1/2} \sum\limits_{k<\Lambda} \varphi_{\bf k} \exp(i {\bf kx})$.
Moreover, the only allowed configurations of $\varphi({\bf x})$ are those, for which
$\varphi_{\bf k}=0$ holds at $k \equiv \mid {\bf k} \mid > \Lambda$. 
This is the limiting case $m \to \infty$ of the
model where all configurations are allowed, but Hamiltonian~(\ref{eq:Hf})
is completed by the term $\sum_{\bf k} \left( k/\Lambda \right)^{2m}
\mid \varphi_{{\bf k}} \mid^2$. 

We consider also the lattice version of the scalar $\varphi^4$ model with the Hamiltonian
\begin{equation}
\frac{\mathcal{H}}{k_B T}  = - \beta \sum\limits_{\langle i j \rangle}
\varphi({\bf x}_i) \varphi({\bf x}_j) + \sum\limits_i \left( \varphi^2({\bf x}_i) + \lambda \left( \varphi^2({\bf x}_i) -1 \right)^2 \right) \;,
\label{eq:lattice}
\end{equation}
where $-\infty < \varphi({\bf x}_i) < \infty$ is a continuous scalar order parameter at the $i$-th lattice site with coordinate ${\bf x}_i$, 
and $\langle ij \rangle$ denotes the set of all nearest neighbors.

The considered here versions of the $\varphi^4$ model are standard and widely used  
in numerous analytical and numerical studies -- see, e.~g. \cite{Amit,Ma,Justin,Kleinert,PV,MHB86,TC90,MF92,Hasenbusch} and references therein.

\section{The critical two--point correlation functions above the \\ upper critical dimension}
\label{Sec:4d}

It is well known~\cite{Amit,Ma,Justin,Kleinert,PV} that the critical behavior of the $\varphi^4$ model~(\ref{eq:H}) above the upper critical spatial
dimension $d$, i.~e., at $d>4$, is determined by the Gaussian fixed point $r_0=0$ and $u=0$. 
It means that the critical correlations functions at $\mid {\bf x} \mid \to \infty$ can be 
exactly calculated from the Gaussian model with Hamiltonian
\begin{equation} \label{eq:HG}
\frac{\cal{H}}{k_B T}= \int c \, (\nabla \varphi)^2 d {\bf x} = \sum\limits_{{\bf k}, \, k <\Lambda} c \,{\bf k}^2 
{\mid \varphi_{\bf k} \mid}^2 \;.
\end{equation}

In the Gaussian model, the correlation functions $\langle \varphi^n({\bf x}_1) \varphi^m({\bf x}_2) \rangle$
can be exactly calculated by coupling the variables according to the Wick's theorem.
The calculations can be performed either for Fourier transforms or directly for the real--space correlation functions.
In the latter case, one has to consider diagrams, constructed from one vertex with $n$ lines at coordinate ${\bf x}_1$
and one vertex with $m$ lines at coordinate ${\bf x}_2$. The correlation function 
$\langle \varphi^n({\bf x}_1) \varphi^m({\bf x}_2) \rangle$ is given by the sum of all diagrams, obtained  
by coupling the lines. In this diagram technique, a line starting at coordinate ${\bf x}_1$ and ending at coordinate
${\bf x}_2$ gives the factor $\langle \varphi({\bf x}_1) \varphi({\bf x}_2) \rangle$. A line can start and end
at the same point, in which case it gives the factor $\langle \varphi^2({\bf x}) \rangle \equiv \langle \varphi^2 \rangle$
for any ${\bf x}$. Taking into account the combinatorial factors for different couplings, we obtain
\begin{equation}
\langle \varphi^n({\bf 0}) \varphi^m({\bf x}) \rangle = 
\sum\limits_{\ell} \frac{n! \, m!}{\ell ! \, [(n-\ell)/2]! \, [(m-\ell)/2]!}
\left( \frac{\langle \varphi^2 \rangle}{2} \right)^{(n+m-2 \ell)/2} 
\langle \varphi({\bf 0}) \varphi({\bf x}) \rangle^{\ell} \;,
\label{eq:corrG}
\end{equation}
where $\ell$ is an odd integer within $1 \le \ell \le \min \{n,m \}$ for odd $n$ and $m$.
Similarly, $\ell$ is an even integer within $0 \le \ell \le \min \{n,m \}$ for even $n$ and $m$.
If $n$ is odd and $m$ is even or vice versa, then the coupling of all lines is not possible, and the correlation
function is zero.

The Fourier--transformed two--point correlation function of the Gaussian model~(\ref{eq:HG}) is well known
to be $G({\bf k})= \langle \mid \varphi_{\bf k} \mid^2 \rangle = 1/(2ck^2)$. It means that
$\langle \varphi({\bf 0}) \varphi({\bf x}) \rangle \propto x^{2-d}$ holds at $x= \mid {\bf x} \mid \to \infty$.
Thus, we conclude from Eq.~(\ref{eq:corrG}) that 
$\langle \varphi^n({\bf 0}) \varphi^m({\bf x}) \rangle$ is proportional to $\langle \varphi({\bf 0}) \varphi({\bf x}) \rangle$
at $x \to \infty$ for positive odd integers $n$ and $m$, whereas the asymptotic behavior is
$\propto (const + \langle \varphi({\bf 0}) \varphi({\bf x}) \rangle^2)$ for positive even integers $n$ and $m$.  
In particular, 
\begin{equation}
 \langle \varphi^n({\bf 0}) \varphi^m({\bf x}) \rangle =  n!! \, m!! \, \langle \varphi^2 \rangle^{(n+m-2)/2} \, 
 \langle \varphi({\bf 0}) \varphi({\bf x}) \rangle + O \left( x^{6-3d} \right)  
\end{equation}
holds for any positive odd integers $n$ and $m$ at $x \to \infty$, where $n!! = 1 \cdot 3 \cdot 5 \cdots n$,
except that the remainder term $O \left( x^{6-3d} \right)$ exactly vanishes for $n=1$ and/or $m=1$.

\section{The two--point joint probability density and \\ correlation functions}
\label{sec:jpd}

All correlation functions $\langle \varphi^n({\bf 0}) \varphi^m({\bf x}) \rangle$ in the $\varphi^4$ 
model~(\ref{eq:H}) (and not only in such a model) can be calculated, if we know the two--point joint probability density
$\mathcal{P}_2(\varphi_1,\varphi_2,{\bf x}_1,{\bf x}_2)$. In this notation, 
$\mathcal{P}_2(\varphi_1,\varphi_2,{\bf x}_1,{\bf x}_2) d \varphi_1 d \varphi_2$ is the probability that
$\varphi({\bf x}_1) \in [\varphi_1,\varphi_1+d \varphi_1]$ and $\varphi({\bf x}_2) \in [\varphi_2,\varphi_2+d \varphi_2]$
at $d \varphi_1 \to 0$ and  $d \varphi_2 \to 0$. Consequently, we have
\begin{equation}
 \langle \varphi^n({\bf x}_1) \varphi^m({\bf x}_2) \rangle
 = \int\limits_{-\infty}^{\infty} \int\limits_{-\infty}^{\infty} \varphi_1^n \varphi_2^m \,
 \mathcal{P}_2(\varphi_1,\varphi_2,{\bf x}_1,{\bf x}_2) \, d\varphi_1 d\varphi_2 \;.
\label{eq:corP}
 \end{equation}

It is convenient to represent the two--point joint probability density as
\begin{equation}
 \mathcal{P}_2(\varphi_1,\varphi_2,{\bf x}_1,{\bf x}_2) = \mathcal{P}_1(\varphi_1) \mathcal{P}_1(\varphi_2)
 + \mathcal{F}(\varphi_1,\varphi_2,\mid {\bf x}_1-{\bf x}_2 \mid) \;,
\label{eq:P}
\end{equation}
where $\mathcal{P}_1(\varphi)$ is the one--point probability density ($\mathcal{P}_1(\varphi) d\varphi$
is the probability that $\varphi({\bf x}) \in [\varphi,\varphi + d\varphi]$ at $d \varphi \to 0$ for any coordinate ${\bf x}$).
The spatial dependence is represented in terms of 
$\mid {\bf x}_1-{\bf x}_2 \mid$ owing to the translational and rotational symmetry of the model.
Moreover, we have $\mathcal{F}(\varphi_1,\varphi_2,x) \to 0$ at $x \to \infty$,
since the correlations vanish at infinitely large distances. Inserting~(\ref{eq:P}) into~(\ref{eq:corP})
at ${\bf x}_1={\bf 0}$ and ${\bf x}_2={\bf x}$, we obtain
\begin{equation}
 \langle \varphi^n({\bf 0}) \varphi^m({\bf x}) \rangle - \langle \varphi^n \rangle \langle \varphi^m \rangle 
 = \int\limits_{-\infty}^{\infty} \int\limits_{-\infty}^{\infty} \varphi_1^n \varphi_2^m \,
 \mathcal{F}(\varphi_1,\varphi_2,\mid {\bf x} \mid) \, d\varphi_1 d\varphi_2 \;.
\label{eq:corF}
 \end{equation}
According to the symmetry of the model, we have $\mathcal{P}_1(\varphi) = \mathcal{P}_1(-\varphi)$,
$\mathcal{P}_2(\varphi_1,\varphi_2,{\bf x}_1,{\bf x}_2)=\mathcal{P}_2(-\varphi_1,-\varphi_2,{\bf x}_1,{\bf x}_2)$
and  $\mathcal{P}_2(\varphi_1,\varphi_2,{\bf x}_1,{\bf x}_2)=\mathcal{P}_2(\varphi_2,\varphi_1,{\bf x}_1,{\bf x}_2)$.
Eq.~(\ref{eq:P}) then implies $\mathcal{F}(\varphi_1,\varphi_2,x)=\mathcal{F}(-\varphi_1,-\varphi_2,x)$ and 
$\mathcal{F}(\varphi_1,\varphi_2,x)=\mathcal{F}(\varphi_2,\varphi_1,x)$.

We need some more specific properties of $\mathcal{F}(\varphi_1,\varphi_2,x)$ for our analysis.
The above relations imply that $\mathcal{F}(0,\varphi_2,x)=\mathcal{F}(0,-\varphi_2,x)$ holds.
This symmetry with respect to the argument $\varphi_2$ does not hold for a fixed $\varphi_1 \ne 0$.
It is quite natural to assume that this symmetry is broken in such a way that
\begin{eqnarray}
 \mathcal{F}(\varphi_1,\varphi_2,x) > \mathcal{F}(\varphi_1,-\varphi_2,x) \qquad \mbox{for } \varphi_1>0 \; \mbox{and } \varphi_2>0 
\label{eq:condi}
 \end{eqnarray}
holds due to the interactions of ferromagnetic type in the $\varphi^4$ model. Indeed, if a spin at a given coordinate is oriented up
($\varphi_1 >0$) then it is energetically preferable and more probable that another spin at any finite distance $x$ from it
is also oriented up, in such a way that the inequality~(\ref{eq:condi}) is satisfied. 

According to the current knowledge about the critical phenomena, the two--point correlation functions are representable by an expansion
in powers of $x$ at the critical point, and sometimes also the logarithmic corrections appear. Therefore, it is natural to assume
that $\mathcal{F}(\varphi_1,\varphi_2,x)$ can be expanded as
\begin{equation}
 \mathcal{F}(\varphi_1,\varphi_2,x) = \sum\limits_{(i,j) \in \Omega} A_{ij}(\varphi_1,\varphi_2) \, x^{-\lambda_i} (\ln x)^{\mu_{ij}} 
\label{eq:expansion}
 \end{equation}
at $x \to \infty$, where $A_{ij}$ are the expansion coefficients, and the sum runs over pairs of indices 
$(i,j)$ belong to some set $\Omega \subset \mathbb{N}^2$. 

Let us $i=j=0$ correspond to the leading term.
The condition~(\ref{eq:condi}) is satisfied at $x \to \infty$, if it is satisfied for this leading expansion term. 
Formally, it is also satisfied in this limit if the asymmetry, stated in~(\ref{eq:condi}),
first shows up in some term with indices $i^*$ and $j^*$, whereas all lower--order terms are symmetric, implying
that $A_{ij}(\varphi_1,\varphi_2) = A_{ij}(\varphi_1,-\varphi_2)=A_{ij}(-\varphi_1,\varphi_2)$ holds for these terms.
The case, where the leading term is asymmetric is also included, and it corresponds to $i^*=j^*=0$.
Consider now odd $n$ and $m$.
In this case, inserting~(\ref{eq:expansion}) into~(\ref{eq:corF}), the symmetric terms give vanishing result,
whereas the leading asymptotic behavior is provided by the term with indices $i^*$ and $j^*$, for which we have 
\begin{eqnarray}
 A_{i^*j^*}(\varphi_1,\varphi_2) > A_{i^*j^*}(\varphi_1,-\varphi_2) \qquad \mbox{for } \varphi_1>0 \; \mbox{and } \varphi_2>0 
\label{eq:condia}
 \end{eqnarray} 
according to~(\ref{eq:condi}). Thus, we obtain
\begin{equation}
 \langle \varphi^n({\bf 0}) \varphi^m({\bf x}) \rangle = B_{nm} \, x^{-\lambda_{i^*}} (\ln x)^{\mu_{i^*j^*}}
 \qquad \mbox{at} \quad x \to \infty
\end{equation}
for odd $n$ and $m$, where $B_{nm}$ are positive coefficients given by
\begin{eqnarray}
 B_{nm} &=& \int\limits_{-\infty}^{\infty} \int\limits_{-\infty}^{\infty} \varphi_1^n \varphi_2^m
 A_{i^*j^*}(\varphi_1,\varphi_2) d\varphi_1 d\varphi_2
\nonumber \\
 &=& 2 \int\limits_{0}^{\infty} \varphi_1^n d\varphi_1 \int\limits_{0}^{\infty} \varphi_2^m
 \left[ A_{i^*j^*}(\varphi_1,\varphi_2) -A_{i^*j^*}(\varphi_1,-\varphi_2 \right)] d\varphi_2 >0 \;.
\label{eq:Bnm}
 \end{eqnarray}
The positiveness of $B_{nm}$ follows from~(\ref{eq:condia}) and implies that
\begin{equation}
 \langle \varphi^n({\bf 0}) \varphi^m({\bf x}) \rangle = C_{nm} \,
 \langle \varphi({\bf 0}) \varphi({\bf x}) \rangle
  \qquad \mbox{at} \quad x \to \infty
\label{eq:propto}
\end{equation}
holds with $C_{nm} = B_{nm}/B_{11}>0$ for any positive odd integers $n$ and $m$. This result is a generalization of those in
Sec.~\ref{Sec:4d} to any spatial dimension, at which the critical point of the second-order phase
transition exists.
 
 The actual consideration can be trivially extended to the lattice version of the scalar $\varphi^4$ model. Indeed, all relations of
this section can be applied to the correlation functions  $\langle \varphi^n({\bf 0}) \varphi^m({\bf x}) \rangle$
with ${\bf x}$ oriented along any given crystallographic direction in the lattice. Only the coefficients 
can depend on the particular lattice and crystallographic direction. Moreover, the two--point correlation functions
are isotropic at large distances $x \to \infty$, so that the asymptotic relation~(\ref{eq:propto})
refers also to the lattice model, $C_{nm}$ being isotropic.
 
\section{Results of Monte Carlo simulation}
\label{sec:MC}
 
\begin{figure}
\begin{center}
\includegraphics[width=0.7\textwidth]{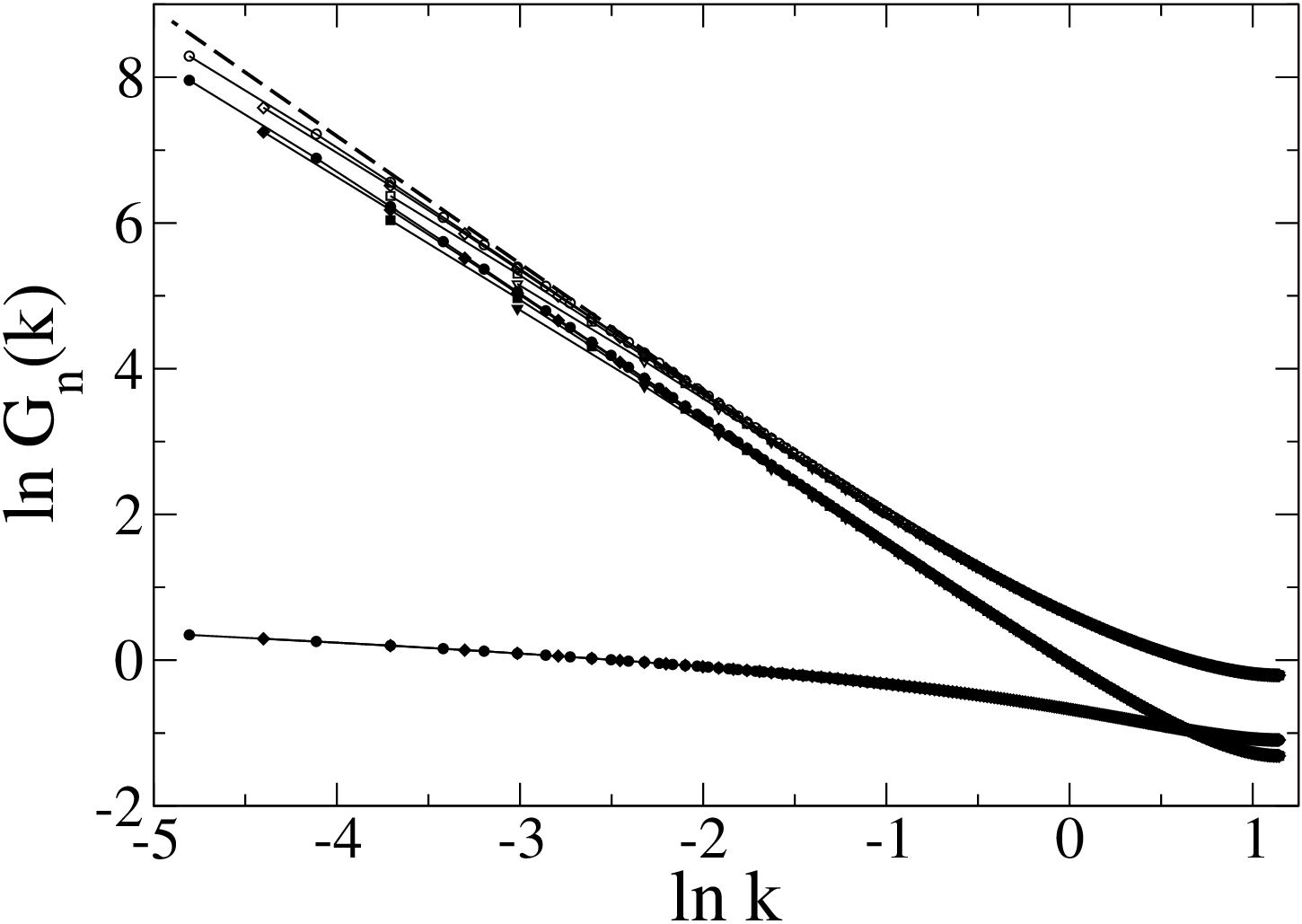}
\end{center}
\caption{The $\ln G_n({\bf k})$ vs $\ln k$ plots at $n=1$ (solid symbols, upper plots),
$n=2$ (solid symbols, lower plots) and $n=3$ (empty symbols, upper plots) for different lattice sizes --
$L=128$ (triangles), $L=256$ (squares), $L=512$ (diamonds) and $L=768$ (circles).
The straight dashed line indicates the small--$k$ asymptotic behavior of $\ln G_3({\bf k})$ 
at $L \to \infty$.}
\label{Fig1}
\end{figure}

We have performed Monte Carlo (MC) simulations for the lattice $\varphi^4$ model~(\ref{eq:lattice})
in two dimensions. Let us denote by $G_n({\bf k})$ the Fourier transform of the two--point correlation function 
$\langle \varphi^n({\bf 0}) \varphi^n({\bf x}) \rangle$. It is calculated as 
$G_n({\bf k}) = \langle \mid v_n({\bf k}) \mid^2 \rangle$, where $v_n({\bf k})$
is the Fourier transform of $\tilde{v}_n({\bf x})=\varphi^n({\bf x})$, i.~e.,
$v_n({\bf k})=N^{-1/2} \sum_{\bf x} \tilde{v}_n({\bf x}) \exp(-i{\bf kx})$, where $N=L^2$
is the number of sites in the 2D square lattice with dimensions $L \times L$. 
The MC simulations at Hamiltonian parameters $\lambda=1$ and $\beta=0.680605$ have been performed,
where the chosen value of $\beta$ is consistent with the critical value $\beta_c=0.680605 \pm 0.000004$
estimated in~\cite{KMRphi4}. The Fourier transforms $G_n({\bf k})$ in the $\langle 10 \rangle$ crystallographic
direction have been evaluated at $n=1,2,3$ for different lattice sizes $L=128, 256, 512$ and $768$, 
using the techniques described in~\cite{KMRphi4} with the only generalization $\varphi({\bf x}) \to \tilde{v}_n({\bf x})$ and
$\varphi_{\bf k} \to v_n({\bf k})$. The obtained $\ln G_n({\bf k})$ vs $\ln k$ plots are shown in Fig.~\ref{Fig1}.

Although the simulations have been performed at an approximate value of $\beta_c$, our techniques allow us to 
recalculate the data for slightly different $\beta$, using the Taylor series. Thus we have verified that the error
$\pm 0.000004$ in the $\beta_c$ value is insignificant, i.~e., it causes systematic
deviations in the plots of Fig.~\ref{Fig1}, which are much smaller than the symbol size.
We can see from these plots that the behavior of $G_3({\bf k})$ is very similar to that of $G_1({\bf k})$
at small enough $k$ values, i.~e., the corresponding log--log plots are practically parallel to each other
for a given lattice size $L$. Some finite--size effects are observed, in such a way that these plots 
converge to a curve with asymptotic slope $-7/4$ at $L \to \infty$, as indicated by a dashed line in Fig.~\ref{Fig1}.
It is consistent with the well known $G_1({\bf k}) \propto k^{-2+\eta}= k^{-7/4}$ asymptotic critical behavior 
in the models of 2D Ising universality class. 

In fact, the MC data suggest that $G_3({\bf k}) \propto G_1({\bf k})$ holds at 
$k \to 0$, which corresponds to $\langle \varphi^3({\bf 0}) \varphi^3({\bf x}) \rangle 
\propto \langle \varphi({\bf 0}) \varphi({\bf x}) \rangle$ at $x \to \infty$,
in agreement with~(\ref{eq:propto}).
To the contrary, the slope of $\ln G_2({\bf k})$ vs $\ln k$ plot is much smaller 
 than that of $\ln G_1({\bf k})$ vs $\ln k$ plot for small $k$ values, indicating that 
 $\langle \varphi^2({\bf 0}) \varphi^2({\bf x}) \rangle - \langle \varphi^2 \rangle^2$ decays much faster than 
 $\langle \varphi({\bf 0}) \varphi({\bf x}) \rangle$ at $x \to \infty$
 and the proportionality relation~(\ref{eq:propto}) does not hold for even exponents $n$ and $m$.

\section{Some results of the conformal field theory}
\label{sec:CFT}

In this section we will briefly review and critically discuss recent results of the conformal field
theory (CFT) reported in~\cite{RT2015}. A method is proposed in this work to recover the $\epsilon$--expansion
from CFT. The operators $V_n \equiv (\varphi^n)_{\mathrm{WF}}$ are considered in the critical $\varphi^4$ theory,
where $(\varphi^n)_{\mathrm{WF}}$ is the operator $\varphi^n$ at the Wilson--Fisher (WF) fixed point 
in $d=4 - \epsilon$ dimensions ($\epsilon >0$). The notation $V_n$ is used to distinguish the operators at the 
WF fixed point from the free theory operators $\varphi^n$. As claimed in~\cite{RT2015},
all operators $\varphi^n$ of the free theory are primaries, whereas
$V_3$ is not a primary, since it is related to $V_1$ via certain equation of motion.

The following axioms are formulated in~\cite{RT2015} as cornerstones of the proposed approach:
\begin{enumerate}
 \item The WF fixed point is conformally invariant. 
 \item Correlation functions of operators at the WF fixed point approach free theory correlators 
 in the limit $\epsilon \to 0$.
 \item Operators $V_n$, $n \ne 3$, are primaries. Operator $V_3$ is not a primary but is
 proportional to $\partial^2 V_1$:
 \begin{equation}
  \partial^2 V_1 = \alpha V_3 \;.
  \label{eq:motion}
 \end{equation}
\end{enumerate}
The coefficient $\alpha$ is unknown at this stage, but later it is found by fitting  
the correlator $\langle V_3({\bf 0}) V_3({\bf x}) \rangle$ to the free theory correlator 
$\langle \varphi^3({\bf 0}) \varphi^3({\bf x}) \rangle$ in four dimensions at $\epsilon \to 0$.

The two--point correlation functions in this CFT behave as
\begin{equation}
 \langle V_n({\bf 0}) V_m({\bf x}) \rangle = B_{nm} \, x^{-\Delta_n - \Delta_m}
\label{eq:corCFT}
 \end{equation}
at the WF fixed point, where $B_{nm}$ are coefficients (which are zero in a subset of cases),
whereas $\Delta_n$ is the dimension of operator $V_n$, which can be expressed as
\begin{equation}
 \Delta_n = n \delta + \gamma_n \;,
\end{equation}
where $\delta = 1- \epsilon/2$ is the free scalar dimension and $\gamma_n$ is the anomalous
dimension of operator $V_n$. Moreover, $\gamma_n \to 0$ holds at $\epsilon \to 0$.

Eq.~(\ref{eq:corCFT}) represents a clearly different scaling form than~(\ref{eq:propto}).
A resolution of this puzzle is such that~(\ref{eq:corCFT}) refers to a formal treatment of
a different $\varphi^4$ model which, contrary to our case, does not contain any upper
cut--off for wave vectors. It formally ensures that the scaling can be
exactly scale--free, i.~e., described exactly by a power law. Note that the upper cut--off
($k < \Lambda$) gives certain length scale $1/\Lambda$, so that the real--space correlation
functions appear to be only asymptotically (at $x \to \infty$) scale--free at the critical point and even at the 
renormalization
group (RG) fixed point in a model with finite $\Lambda$. Indeed, Eq.~(\ref{eq:corCFT}) cannot hold
at $x \to 0$ in such a model, since finite $\Lambda$ ensures that the local 
quantities $\langle V_n({\bf 0}) V_m({\bf 0}) \rangle$ are
also finite. 

Note that an exact power law is observed in the usual Wilson--Fisher renormalization
in the wave--vector space, however, only for the Fourier--transformed two--point correlation
function $G({\bf k})$. It behaves as $G({\bf k})=ak^{-2+\eta}$ within $0<k<\Lambda$ 
at the Wilson--Fisher fixed point owing to the well--known rescaling rule 
$G({\bf k},\mu) = s^{2-\eta}G(s{\bf k},R_s \mu)$~\cite{Ma}. Here $\mu$ is a point in the parameter space
of the Hamiltonian. This point is varied under the RG transformation $R_s$. The real--space correlation 
function $\tilde{G}({\bf x})$ can be straightforwardly calculated from $G({\bf k})=ak^{-2+\eta}$ via 
$\tilde{G}({\bf x}) = (2 \pi)^{-d} \int_{k<\Lambda} G({\bf k}) \exp(i{\bf kx}) d^dk$
to aware that it does not follow an exact power law.

Apparently, the absence of any length scale
is a desired property in CFT. Unfortunately, the model without any cut--off is ill--defined,
since the local quantities $\langle V_{2n}({\bf x}) \rangle$ ($n \ge 1$) are divergent. 
Moreover, no mathematically justified calculations can be performed in this case even in the Gaussian model
because of the divergent quantity $\langle \varphi^2 \rangle$ appearing in~(\ref{eq:corrG}).
Recall that a ``tadpole'' \tadpole is formed when two lines of
the same vertex are coupled, and it gives the factor $\langle \varphi^2 \rangle$.
Obviously, the free theory calculations at $d=4$ are performed in~\cite{RT2015}, omitting
the divergent diagrams with tadpoles in the Gaussian model without any cut--off. 
Only in this case we can obtain the reported in~\cite{RT2015} results, $\langle \varphi({\bf 0}) \varphi({\bf x}) \rangle = x^{-2}$
and $\langle \varphi^3({\bf 0}) \varphi^3({\bf x}) \rangle = x^{-6}$, after certain normalization
of $\varphi$. 
Such a subtraction of divergent terms is, in fact, a well known idea of the 
perturbative renormalization. From this point of view,
the approach of~\cite{RT2015} is neither rigorous nor really non--perturbative.

We think that the only way how this formal CFT can be justified is
to find a strict relation between this formal treatment and the results of a well
defined model. Such possible relations are discussed in Sec.~\ref{sec:pertest}.

Another question is the validity of the ``equation of motion''~(\ref{eq:motion}).
In the $\varphi^4$ model, each configuration $\{\varphi \}$ of the
order parameter field $\varphi({\bf x})$ shows up with the statistical weight 
$Z^{-1} \exp[-\mathcal{H}(\{ \varphi \})/(k_BT)]$, where $Z$ is the partition function. 
Therefore, it is impossible that a constraint
of the form~(\ref{eq:motion}) is really satisfied, unless any local violation of such a constraint
leads to a divergent local Hamiltonian density $h({\bf x})$.
 To aware about this, we can consider a configuration $\{\varphi \}$, which obeys~(\ref{eq:motion}), and
another configuration $\{\varphi \}'$. Let us $\{\varphi \}'=\{\varphi \}$ holds 
everywhere, except a local region ${\bf x} \in \Omega$, where $\{\varphi \}'$ does not satisfy~(\ref{eq:motion}). 
The ratio of statistical weights for $\{\varphi \}'$ and $\{\varphi \}$ is 
$Z^{-1} \exp[-\Delta \mathcal{H}(\{ \varphi \})/(k_BT)]$, where
$\Delta \mathcal{H}= \int_{{\bf x} \in \Omega} (h'({\bf x})-h({\bf x})) d{\bf x}$, $h({\bf x})$
and $h'({\bf x})$ being Hamiltonian densities for configurations $\{\varphi \}$
and $\{\varphi \}'$, respectively. If~(\ref{eq:motion}) really holds, then the configuration
$\{\varphi \}'$ must show up with a vanishing statistical weight as compared to that of $\{\varphi \}$,
i.~e., the above mentioned ratio must be zero. Obviously, it is possible only if $h'({\bf x})$ is
divergent at least within a subregion of $\Omega$. 

On the other hand, there are no evidences that the density of renormalized Hamiltonian contains any term 
which is divergent if~(\ref{eq:motion}) is violated. Particularly, the usual perturbative approximations for the fixed--point
Hamiltonian in $4 - \epsilon$ dimensions suggest that the renormalized Hamiltonian is just the same as the bare one~(\ref{eq:H})
with the only difference that the Hamiltonian parameters have special, i.~e., renormalized values.
The density of such renormalized Hamiltonian is finite for relevant field configurations with 
$\mid \nabla \varphi({\bf x}) \mid < \infty$ and $\mid \varphi({\bf x}) \mid < \infty$.

\section{Possible relations of formal CFT to well defined models}
\label{sec:pertest}
 
 A standard way to compare the results of the lattice $\varphi^4$ model to those of CFT is to
redefine operators of the lattice model. In particular, $\varphi^3$ has to be replaced by
$W_3=\varphi^3 + b \varphi$, where an appropriate value of the constant $b$ is chosen to cancel 
the $\langle \varphi({\bf 0}) \varphi({\bf x}) \rangle$ contribution contained in 
$\langle \varphi^3({\bf 0}) \varphi^3({\bf x}) \rangle$. In this case, the operator $\varphi \equiv W_1$ remains
unaltered. We have verified perturbatively
that the same method works also in the considered here continuous $\varphi^4$ model with cut--off for
wave vectors. Namely, the perturbative expansion in powers of $x^{-2}$, $\ln x$ and $\epsilon$
is such that the terms with $x^{-2}$ (and logarithmic correction factors) cancel in 
$\langle W_1({\bf 0}) W_3({\bf x}) \rangle$ and the terms with both $x^{-2}$ and $x^{-4}$
cancel in $\langle W_3({\bf 0}) W_3({\bf x}) \rangle$ at the RG fixed point,
if $b$ is chosen as $b=-C_{13}$, where $C_{13}$ is the coefficient in~(\ref{eq:propto}).
We have examined the corresponding expansion in powers of $k^2$, $\ln k$ and $\epsilon$
for the Fourier transforms of these correlators, using the standard representation of
$G({\bf k})$ by self-energy diagrams, to verify that the above mentioned cancellations
take place in all orders of the perturbation theory. It ensures that the expected
$\propto x^{-\Delta_1-\Delta_3}$ asymptotic behavior of $\langle W_1({\bf 0}) W_3({\bf x}) \rangle$
and the expected $\propto x^{-2 \Delta_3}$ asymptotic behavior of $\langle W_3({\bf 0}) W_3({\bf x}) \rangle$
can be, in principle, consistent with the perturbation theory, since
$\Delta_n \to n$ at $\epsilon \to 0$.

We have tested also two--point correlators of odd operators $W_1 \equiv \varphi$,
$W_3=\varphi^3 + b \varphi$ and $W_5=\varphi^5 + b_1 \varphi + b_3 \varphi^3$.
Unfortunately, this method becomes problematic if operators $W_{2m+1}$ with $m>1$ are included,
since the number of required cancellations increase more rapidly than the number of free
coefficients. Note that terms with $x^{-2}$ and $x^{-4}$ have to be canceled 
in $\langle W_1({\bf 0}) W_5({\bf x}) \rangle$, terms with $x^{-2}$, $x^{-4}$ and $x^{-6}$
have to be canceled in $\langle W_3({\bf 0}) W_5({\bf x}) \rangle$, and 
terms with $x^{-2}$, $x^{-4}$, $x^{-6}$ and $x^{-8}$
have to be canceled in $\langle W_5({\bf 0}) W_5({\bf x}) \rangle$.
We have found it already impossible to reach all the necessary cancellations
up to the $O \left(\epsilon^2 \right)$ order for all two--point correlators
of the above mentioned three operators $W_1$, $W_3$ and $W_5$.
Hence, the finding of a precise relation between a well--defined continuous $\varphi^4$ model
and the formal CFT treatment of~\cite{RT2015} is at least problematic within the perturbation
theory. The details of our perturbative calculations are given in Appendix.

It is interesting to look for a possible non--perturbative relation between the discussed here 
formal CFT and a well--defined model.
The reported in~\cite{Showk} agreement of the results of the conformal bootstrap method with the known exact spectrum of the
operator dimensions in the 2D Ising model is a good evidence for the existence of such a relation in two dimensions. 

A nontrivial MC evidence for the existence of conformal symmetry in three dimensions has been provided in~\cite{MCconf}.
Another interesting MC test has been performed in~\cite{Janke} to check whether or not relations for amplitudes
of correlation lengths,
analogous to those predicted by CFT in the 2D Ising model, can be found in the 3D Ising model and, more generally,
in the $O(n)$--symmetric models in three dimensions.
Surprisingly, such a relation has been indeed obtained, however, only for anti-periodic boundary conditions and not for
periodic ones, for which this relation exists in two dimensions. It reveals a qualitative difference between the
two--dimensional and three--dimensional cases. 
According to this result, the following judgment can be made:
even if the conformal symmetry exists in the 3D Ising model, then it is, however, questionable
whether or not (or in which sense) the spatial dimensionality $d$ can be considered as a continuous parameter in the CFT. 
On the other hand, $d$ appears as a continuous parameter in the particular treatment of CFT in~\cite{RT2015},
since it is possible to recover the $\epsilon$--expansion from it.

\begin{figure}
\begin{center}
\includegraphics[width=0.48\textwidth]{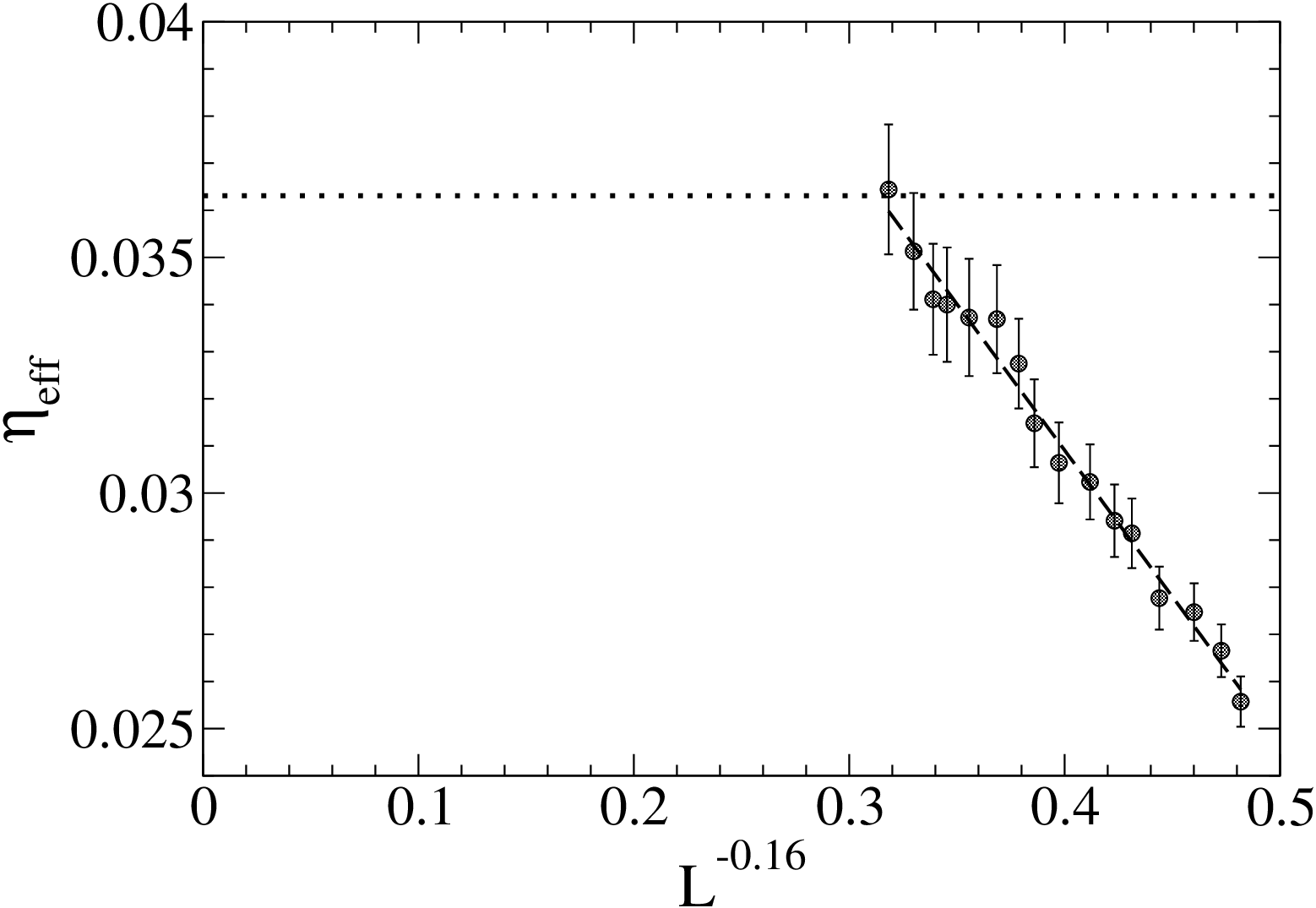} \hfill 
\includegraphics[width=0.48\textwidth]{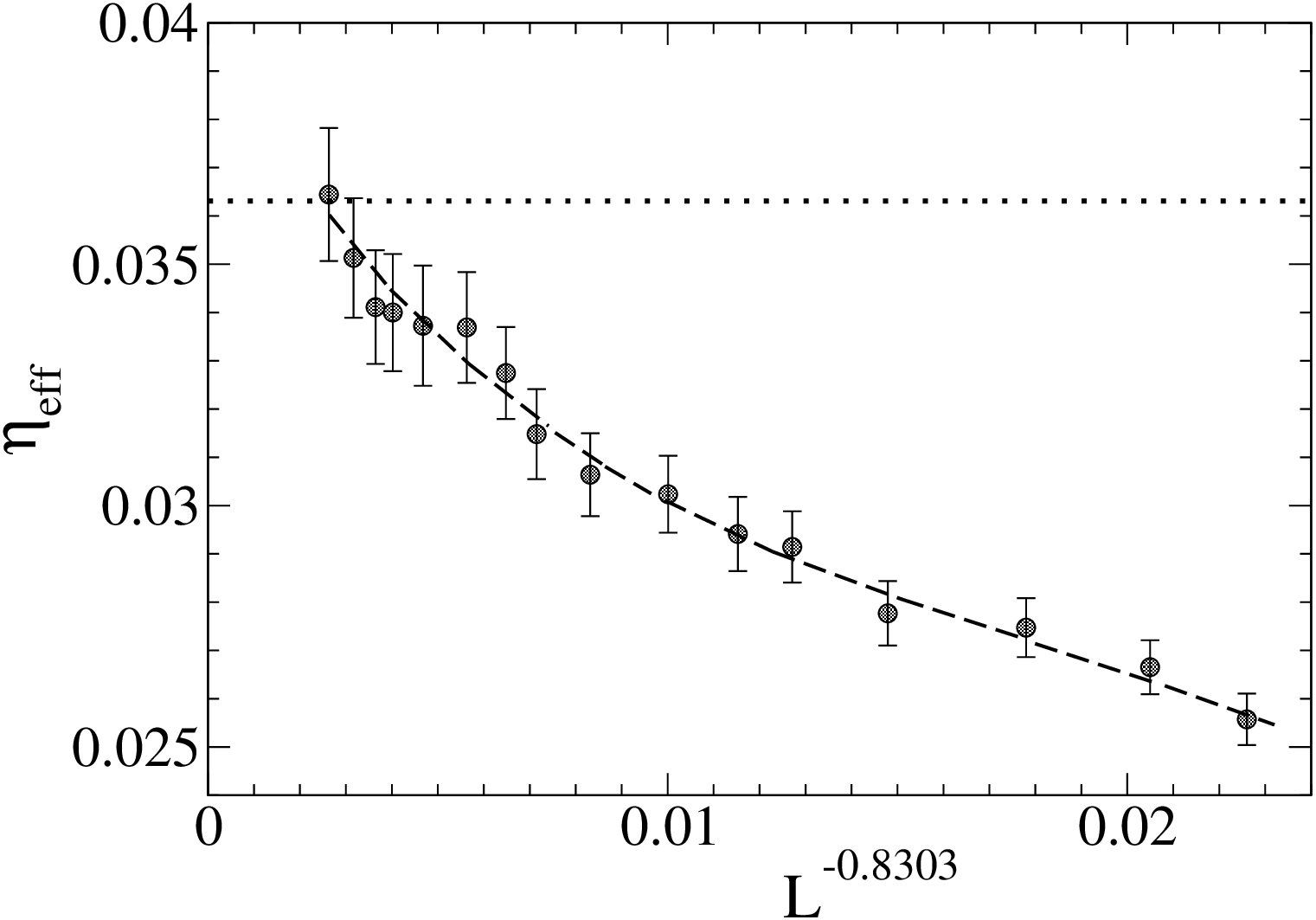}
\end{center}
\caption{The effective exponent $\eta_{\mathrm{eff}}(L)$ depending on $L^{-0.16}$ (left)
and $L^{-0.8303}$ (right). The dashed straight lines represent the linear fit (left)
and a guide to eye (right). The value $\eta = 0.03631(3)$ of~\cite{xx} is indicated by dotted lines.}
\label{fig2}
\end{figure}

Despite of the discussed here problems, the critical exponents provided by this CFT appear to be very well
consistent with the currently accepted values of the 3D Ising model. In particular, $\eta = 0.03631(3)$, $\nu =0.62999(5)$
and $\omega=0.8303(18)$ are reported in~\cite{Showk}, in close agreement with the MC estimates $\eta = 0.03627(10)$,
$\nu = 0.63002(10)$ and $\omega = 0.832(6)$ of Hasenbusch~\cite{Hasenbusch}. These MC results have been obtained
from the data for linear lattice sizes $L \le 360$. We have tested the exponent $\eta$, evaluated from the susceptibility
($\chi$) data for much larger lattice sizes up to $L=2560$. The $\chi/L^2$ data at certain pseudocritical
couplings $\widetilde{\beta}_c(L)$ (tending to the true critical coupling $\beta_c$ at $L \to \infty$ and corresponding to
$U=\langle m^4 \rangle / \langle m^2 \rangle^2=1.6$, where $m$ is the magnetization per spin) for $L \le 1536$ are given in~\cite{KMR_2013}.
Here we add more recent results: $\chi/L^2 = 1.1882(20)$ for $L=1728$, $\chi/L^2 = 1.1741(27)$ for $L=2048$
and $\chi/L^2 = 1.1669(28)$ for $L=2560$, obtained by the same techniques as before. The effective critical
exponent $\eta_{\mathrm{eff}}(L)$ is obtained by fitting the data to the ansatz $\chi/L'^2=aL'^{-\eta}$
within $L' \in [L/2,2L]$. It is expected that $\eta_{\mathrm{eff}}(L) - \eta \propto L^{-\omega}$ holds
at $L \to \infty$, where $\omega$ is the correction--to--scaling exponent. Surprisingly, we observe that 
$\eta_{\mathrm{eff}}$ vs $L^{-\omega}$ behavior shows the best linearity for large lattice sizes at $\omega = 0.16(36)$
rather than at $\omega=0.8303(18)$. The plots of
$\eta_{\mathrm{eff}}(L)$ depending on $L^{-0.16}$ and $L^{-0.8303}$ are shown in Fig.~\ref{fig2}, where the value
$\eta = 0.03631(3)$ of~\cite{Showk} is indicated by dotted lines. 

Although the agreement of this value with currently accepted ones is very good,
it would be not superfluous to make further tests in order to see whether $\eta_{\mathrm{eff}}(L)$
really converges to $\eta = 0.03631(3)$ or, perhaps, to a larger value, as it is suggested
by an extrapolation of plots (dashed lines) in Fig.~\ref{fig2}. It would be also very useful
to obtain a more precise value of $\omega$ from large lattice sizes to see whether or not this value
is indeed remarkably smaller than $0.83$. The latter possibility is strongly supported by our recent
analytical results~\cite{KMRphi4}, from which $\omega \le (\gamma -1)/\nu \approx 0.38$ is expected.

The referred here critical exponents $\eta = 0.03631(3)$, $\nu =0.62999(5)$
and $\omega=0.8303(18)$ have been obtained in~\cite{Showk} based on the following hypotheses: 
\begin{enumerate}
\item[(i)] 
There exists a sharp kink on the border of the two--dimensional region of the allowed values of 
the operator dimensions $\Delta_{\sigma} = (1+\eta)/2$ and $\Delta_{\epsilon} = 3 - 1/\nu$; 
\item[(ii)] 
Critical exponents of the 3D Ising model correspond just to this kink.
\end{enumerate}
If the hypothesis~(i) about the existence of a sharp kink is true, then 
this kink, probably, has a special meaning for the 3D Ising model. Its existence, however, is not evident. 
\begin{figure}
\begin{center}
\includegraphics[width=0.6\textwidth]{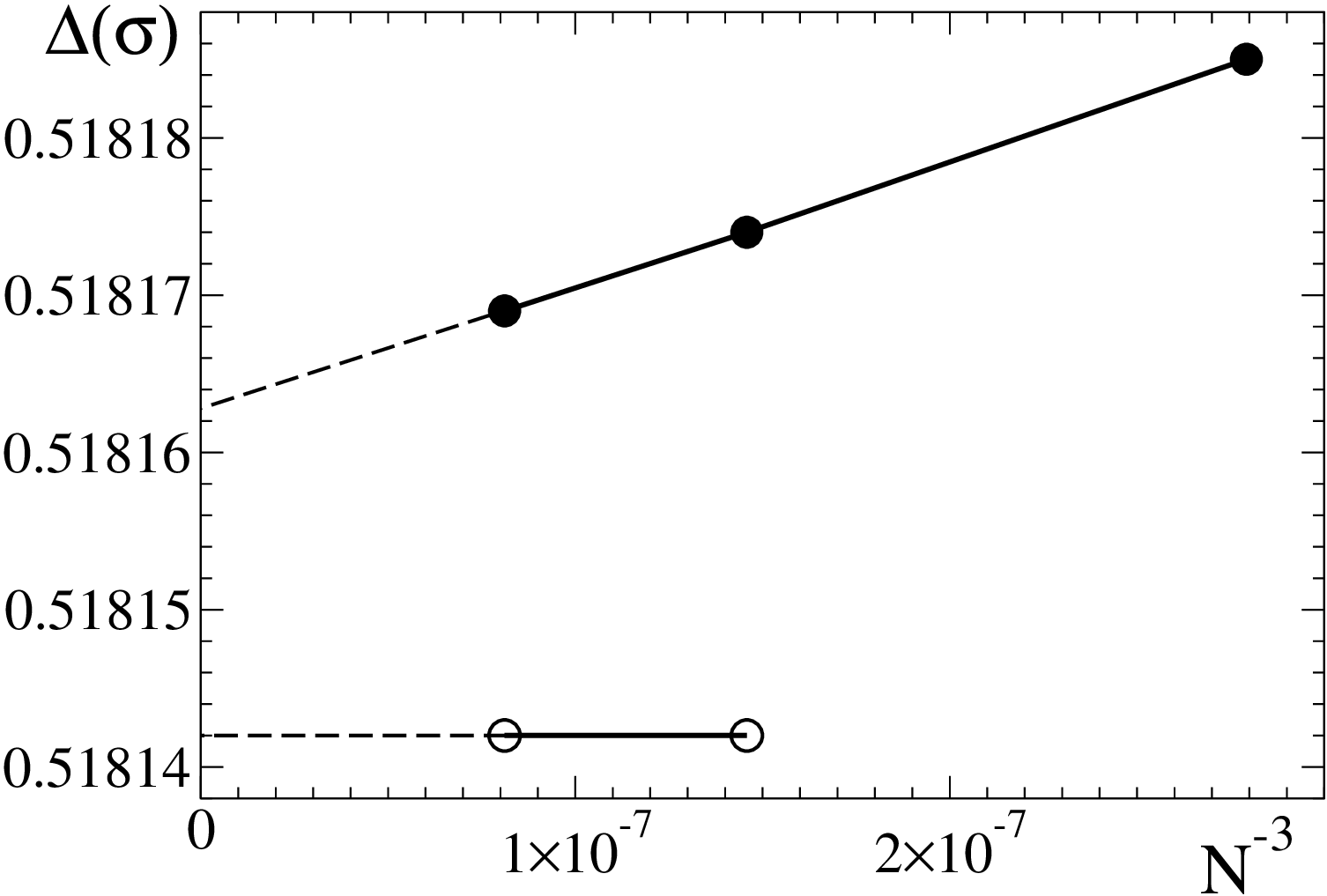}
\end{center}
\caption{The values of $\Delta(\sigma)$, corresponding to the
minimum (solid circles) and the ``kink'' (empty circles) in the plots of Fig.~7 in~\cite{Showk}
depending on $N^{-3}$. The dashed lines show linear extrapolations.}
\label{fig3}
\end{figure}
According to the conjectures of~\cite{Showk}, such a kink is formed at $N \to \infty$, where $N$ is the number of derivatives
included into the analysis. As discussed in~\cite{Showk}, it implies that the minimum in the plots of Fig.~7 
in~\cite{Showk} should be merged with the apparent ``kink''
at $N \to \infty$. This ``kink'' is not really sharp at a finite $N$. 
Nevertheless, its location can be identified with the value of $\Delta(\sigma)$, at which
the second derivative of the plot has a local maximum. The minimum 
of the plot is slightly varied with $N$, whereas the ``kink'' is barely moving~\cite{Showk}.
Apparently, the convergence to a certain asymptotic curve is remarkably faster than $1/N$, as it can be expected 
from Fig.~7  and other similar figures in~\cite{Showk}. In particular, we have found
that the location of the minimum in Fig.~7 of~\cite{Showk} is varied almost linearly with $N^{-3}$.
We have shown it in Fig.~\ref{fig3} by solid circles, the position
of the ``kink'' being indicated by empty circles. The error bars of 
$\pm 0.000001$ correspond to the symbol size. The results for $N=153, 190, 231$ are presented,
skipping the estimate for the location of the ``kink'' at $N=153$, which cannot be well determined
from the corresponding plot in Fig.~7 of~\cite{Showk}.
The linear extrapolations (dashed lines) suggest that the 
minimum, very likely, is moved only slightly closer to the ``kink'' when $N$ is varied from
$N=231$ to $N = \infty$. The linear extrapolation might be too inaccurate. Only in this case  
a refined numerical analysis for larger $N$ values 
can possibly confirm the hypothesis about the formation of a sharp kink at $N \to \infty$.

\section{Summary and conclusions}

We have shown by different non--perturbative methods in Secs.~\ref{Sec:4d} -- \ref{sec:MC} that 
the critical two--point correlation functions 
$\langle \varphi^n({\bf 0}) \varphi^m({\bf x}) \rangle$ are proportional to
$\langle \varphi({\bf 0}) \varphi({\bf x}) \rangle$ at $x \to \infty$
for any positive odd integers $n$ and $m$ in the considered here well--defined continuous and lattice $\varphi^4$ models.
Moreover, this is unambiguously true also in the diagrammatic perturbation theory, as 
shown in the Appendix. This behavior is not consistent with
the form~(\ref{eq:corCFT}) of the conformal field theory (CFT)~\cite{RT2015,Showk,xx}.

The results of CFT in~\cite{RT2015,Showk} are analyzed in Sec.~\ref{sec:CFT}.
In particular, we have found this treatment to be rather formal,
since Eq.~(\ref{eq:corCFT}) apparently 
is obtained by a purely formal treatment of an ill--defined 
$\varphi^4$ model without any cut-off for wave vectors. Moreover, as explained in Sec.~\ref{sec:CFT}, the ``equation of motion''
used in~\cite{RT2015} cannot be expected to hold from the point of view of statistical 
physics. 

According to these problems, we argue that 
the only way how the discussed here formal CFT can be justified is
to find a strict relation between this formal treatment and the results of a well
defined model. This issue has been discussed in Sec.~\ref{sec:pertest}, and some
problems have been revealed, which make the existence of such a relation questionable
for 3D models.

The accurate agreement of the critical exponents of this CFT with
the currently accepted values for the 3D Ising model 
might imply that, despite of the mentioned here problems, this CFT produces correct predictions.

However, there is also a possibility that the correct asymptotic values in reality deviate from those of
this CFT, as suggested by our MC simulation data for very large lattices with linear sizes
up to $L=2560$ -- see Fig.~\ref{fig2} in Sec.~\ref{sec:pertest}.
The plots of effective exponent $\eta_{\mathrm{eff}}$ in this figure increase with $L$ and
stop at an intriguing point just reaching a value, which is very close to the CFT value $\eta = 0.03631(3)$.
Therefore, it would be interesting to obtain some data for even larger lattice sizes to
make a precise statement concerning the convergence (or not convergence) to the value $0.03631(3)$.

\section*{Appendix}

As it is well known~\cite{Ma}, the Fourier transformed two--point correlation function $G({\bf k})$
can be represented perturbatively by the diagram expansion 
\begin{equation}
 G({\bf k}) =  \selfeko  + \selfekk + \selfekkk  + \ldots
\label{Gke}
 \end{equation}
where \selfekk is the perturbative sum of irreducible (self--energy) connected diagrams with the 
wave vectors $\pm {\bf k}$ exiting (or entering) the diagram, and with the Gaussian correlation function
$G_0({\bf k})$ related to the coupling lines.
These diagrams are irreducible in the sense that they
cannot be split in two parts by breaking only one coupling line. Denoting symbolically this 
perturbation sum as \wl, the Fourier transforms of the correlators
$\langle \varphi({\bf 0}) \varphi^3({\bf x}) \rangle$ and $\langle \varphi^3({\bf 0}) \varphi^3({\bf x}) \rangle$,
i.~e., $G_{13}({\bf k})$ and $G_{33}({\bf k})$, can be represented as
\begin{eqnarray}
 G_{13}({\bf k}) &=& 3 \wltad  + \wlDthree = G({\bf k}) \left( 3 D_0+D_3({\bf k}) \right)
 \label{G13k} \;, \\
 G_{33}({\bf k}) &=& 9 \twotad + \DwlD + 6 \tadDthree + \Dtthree \label{G33k} \\
                 &=& G({\bf k}) \left( 3 D_0+D_3({\bf k}) \right)^2 + D_{33}({\bf k}) \;,
\end{eqnarray}
where $D_0$ is the perturbative sum of diagrams of a tadpole \tadpolex , including diagrams
like \tadpole, \tadpoley, etc., where any diagrammatic block is connected by two coupling lines
to the lower node. The combinatorial factor 3 in~(\ref{G13k}) shows up because of three
possibilities to choose the lines of the $\varphi^3$ vertex for the tadpole. 
Quantity $D_3({\bf k})$ is the perturbative sum of all irreducible (in the same sense as before) 
diagrams of the kind \Dthree, where three
coupling lines on the right hand side come from the $\varphi^3$ vertex, 
whereas the line
on the left hand side belongs to one of the vertices of the Hamiltonian, the factor of this line
being canceled. In this symbolic notation, the considered here lines are
connected to any possible inner part of the diagram, summing up over all such possibilities. 
Similarly, the quantity $D_{33}({\bf k})$ contains all irreducible 
diagrams of the kind \Dtthree, where three coupling lines
on both sides come from $\varphi^3$ vertices.

The wave vector ${\bf k}$ is related to the wavy line in the diagrams of~(\ref{G13k})--(\ref{G33k}), 
whereas $\pm {\bf k}$ in these diagrams, as well as in $D_3({\bf k})$ and $D_{33}({\bf k})$, 
is the sum of wave vectors entering the shown by dots external nodes, this sum 
being zero for all internal nodes. The external nodes belong to the vertices, representing the operators
of the considered correlator. These nodes are not shown in~(\ref{Gke}), 
but can be added there. The internal nodes of the summed up diagrams of the symbolically
shown structure belong to the vertices of the Hamiltonian. 

The perturbative sums, represented by the
symbolic diagrams \selfekk, \tadpolex, \Dthree and \Dtthree are formal in the sense that they do not really convergence
without an appropriate resummation. Nevertheless, we can recover
the perturbative expansion up to any desired order by deciphering these symbolic diagrams.
Each specific diagram, contained there, is defined in accordance with the 
generally known rules of the diagram technique. These diagrams are summed up, taking into account
the weight factors of the vertices in the Hamiltonian and the necessary combinatorial factors.

 More generally, the three lines, connected to a node in the considered here diagrams, can come
from a $\varphi^l$ vertex with $l=n$ or $l=m$, if the Fourier transform $G_{nm}({\bf k})$ of the correlator 
$\langle \varphi^n({\bf 0}) \varphi^m({\bf x}) \rangle$ is considered. A useful generalization of $D_3({\bf k})=\Dthree$ is $D_n({\bf k})$, where
three coupling lines on the right hand side are replaced by $n$ coupling lines. Furthermore, the sum of irreducible diagrams of the 
kind \Dtthree, but with $n$ coupling lines on the left hand side and $m$ coupling lines on the right hand side
will be denoted as $D_{nm}({\bf k})$. Let us introduce two extra quantities defined by
\begin{eqnarray}
 B_3({\bf k}) &=& 3 D_0 + D_3({\bf k}) \;, \\
 B_5({\bf k}) &=& 15 D_0^2 + 10 D_0 D_3({\bf k}) + D_5({\bf k}) \;.
\end{eqnarray}
Using these notations, the correlation functions $G_{nm}({\bf k})$ with odd indices $n \le m \le 5$ can be compactly written as
\begin{eqnarray}
 G_{13}({\bf k}) &=& G({\bf k}) B_3({\bf k}) \;, \label{eq:G13} \\
 G_{33}({\bf k}) &=& G({\bf k}) B_3^2({\bf k}) + D_{33}({\bf k}) \;, \\
 G_{15}({\bf k}) &=& G({\bf k}) B_5({\bf k}) \;, \\
 G_{35}({\bf k}) &=& G({\bf k}) B_3({\bf k}) B_5({\bf k}) + 10D_0 D_{33}({\bf k}) + D_{35}({\bf k}) \;, \\
 G_{55}({\bf k}) &=& G({\bf k}) B_5^2({\bf k}) + 100 D_0^2 D_{33}({\bf k}) + 20 D_0 D_{35}({\bf k}) + D_{55}({\bf k}) \label{eq:G55} \;,
\end{eqnarray}
Eqs.~(\ref{eq:G13})--(\ref{eq:G55}) are consistent with~(\ref{eq:propto}), i.~e.,
\begin{equation}
 C_{nm} = B_n({\bf 0}) B_m({\bf 0}) 
\end{equation}
for $n=1,3,5$ and $m=1,3,5$, where $B_1({\bf k}) = 1$. 

In the following, we consider new operators
\begin{eqnarray}
 W_1 &=& \varphi \;, \label{eq:W1} \\
 W_3 &=& \varphi^3 + b \varphi \;,\\
 W_5 &=& \varphi^5 + b_1 \varphi + b_3 \varphi^3 \;, \label{eq:W5}
\end{eqnarray}
where the coefficients have to be chosen such to reach the cancellation of terms of the kind $\propto x^{\gamma} (\ln x)^p$ 
(where proportionality coefficients contain powers of $\epsilon$) with $\gamma >-n-m$ 
and nonnegative integer $p$ in the $\epsilon$--expansion of $\langle W_n({\bf 0}) W_m({\bf x}) \rangle$ at the RG fixed point. It implies
the cancellation of singular (at $k=0$) terms $\propto k^{\gamma'} (\ln k)^p$ with $\gamma'<n+m-4$ in the $\epsilon$--expansion of the corresponding
Fourier transforms $\Gamma_{nm}({\bf k})$. Recall that the expansion of $G({\bf k})$ contains terms of the kind $\propto k^{-2} (\ln k)^p$.
It implies that the leading singularities, which come from $G({\bf k})$, must be canceled in the expansion of $\Gamma_{nm}({\bf k})$
with $n+m >2$. Applying this condition to $\Gamma_{13}({\bf k})$, $\Gamma_{33}({\bf k})$ and $\Gamma_{15}({\bf k})$,
we easily find 
\begin{eqnarray}
 b &=& -C_{13} = -B_3({\bf 0}) \;, \\
 b_1 &=& -b_3 B_3({\bf 0}) - B_5({\bf 0}) \;.
\end{eqnarray}
It yields
\begin{eqnarray}
 \Gamma_{13}({\bf k}) &=& G({\bf k}) \left( B_3({\bf k})-B_3({\bf 0}) \right) \;, \label{eq:Gam13} \\
 \Gamma_{33}({\bf k}) &=& G({\bf k}) \left( B_3({\bf k})-B_3({\bf 0}) \right)^2 + D_{33}({\bf k}) \;, \\
 \Gamma_{15}({\bf k}) &=& G({\bf k}) \left( B_5({\bf k})-B_5({\bf 0}) + b_3 [B_3({\bf k})-B_3({\bf 0})] \right) \;, \label{eq:Gam15} \\
 \Gamma_{35}({\bf k}) &=& G({\bf k}) \left( B_3({\bf k})-B_3({\bf 0}) \right) 
 \left( B_5({\bf k})-B_5({\bf 0})  + b_3 [B_3({\bf k})-B_3({\bf 0})] \right) \nonumber \\
 && \hspace*{2ex} + \left( 10 D_0 + b_3 \right) D_{33}({\bf k}) + D_{35}({\bf k}) \;, \label{eq:Gam35} \\
 \Gamma_{55}({\bf k}) &=& G({\bf k}) \left( B_5({\bf k})-B_5({\bf 0})  + b_3 [B_3({\bf k})-B_3({\bf 0})] \right)^2 \nonumber \\
 && \hspace*{2ex} + \left( 10 D_0 + b_3 \right)^2 D_{33}({\bf k}) + 2 \left( 10D_0+b_3 \right) D_{35}({\bf k}) + D_{55}({\bf k}) \label{eq:Gam55} \;.
 \end{eqnarray}
Eqs.~(\ref{eq:Gam13}) -- (\ref{eq:Gam55}) already ensure the desired properties of $\Gamma_{13}({\bf k})$ and $\Gamma_{33}({\bf k})$,
as well as the cancellation of singularities of the kind $\propto k^{-2} (\ln k)^p$ in all these equations.
The desired cancellation of the $\propto (\ln k)^p$ terms in~(\ref{eq:Gam15}) -- (\ref{eq:Gam55}),
 the $\propto k^2 (\ln k)^p$ terms in~(\ref{eq:Gam35}) -- (\ref{eq:Gam55}) and
the $\propto k^4 (\ln k)^p$ terms in~(\ref{eq:Gam55}) with $p \ge 1$ in all these cases is still not reached.
One can hope to reach this by adjusting the free coefficient $b_3$.

The cancellation of $\propto (\ln k)^p$ terms in~(\ref{eq:Gam15}) implies that the $\propto k^2 (\ln k)^p$
terms have to be canceled in the expression $Q({\bf k}) = B_5({\bf k})-B_5({\bf 0}) + b_3 [B_3({\bf k})-B_3({\bf 0})]$.
If this condition is satisfied, then such terms are canceled also in the expression 
$G({\bf k}) \left( B_3({\bf k})-B_3({\bf 0}) \right) Q({\bf k})$ contained in~(\ref{eq:Gam35}).
The complete cancellation of such terms in~(\ref{eq:Gam35}) is thus reached if the terms of this kind
are canceled in the expression $R({\bf k}) = \left( 10 D_0 + b_3 \right) D_{33}({\bf k}) + D_{35}({\bf k})$.
Furthermore, if the above mentioned cancellation in $Q({\bf k})$ takes place, then the 
terms $\propto k^2(\ln k)^p$ are canceled in the expression 
$G({\bf k}) Q^2({\bf k})$ in the first line of~(\ref{eq:Gam55}).
The expression in the second line of~(\ref{eq:Gam55}) can be written
as $(10D_0+b_3)[R({\bf k}) + D_{35}({\bf k})] + D_{55}({\bf k})$.
Consequently, if the terms $\propto k^2(\ln k)^p$ are canceled in $R({\bf k})$,
then the desired cancellation in~(\ref{eq:Gam55}) is reached at the 
necessary condition that these terms are canceled in $(10D_0+b_3)D_{35}({\bf k}) + D_{55}({\bf k})$.
Moreover, the cancellation must take place at any order of the $\epsilon$--expansion. We consider
the order $O(\epsilon)$ and find out that $D_{55}({\bf k})$ does not contain any term of the kind $\epsilon \, k^2(\ln k)^p$
with $p \ge 1$, whereas $D_{35}({\bf k})$ contains such a term. Thus, we obtain
\begin{equation}
 b_3 = -10 D_0
\end{equation}
as the necessary condition for the discussed here desired cancellations in Eqs.~(\ref{eq:Gam15}) -- (\ref{eq:Gam55}). 
At this condition, Eqs.~(\ref{eq:Gam13}) -- (\ref{eq:Gam55}) are reduced to
\begin{eqnarray}
 \Gamma_{13}({\bf k}) &=& G({\bf k}) (D_3({\bf k})-D_3({\bf 0})) \;, \\
 \Gamma_{33}({\bf k}) &=& G({\bf k}) (D_3({\bf k})-D_3({\bf 0}))^2 + D_{33}({\bf k}) \;, \\
 \Gamma_{15}({\bf k}) &=& G({\bf k}) (D_5({\bf k})-D_5({\bf 0})) \;, \label{eq:Gamm15} \\
 \Gamma_{35}({\bf k}) &=& G({\bf k}) (D_3({\bf k})-D_3({\bf 0}))(D_5({\bf k})-D_5({\bf 0})) + D_{35}({\bf k}) \;, \\
 \Gamma_{55}({\bf k}) &=& G({\bf k}) (D_5({\bf k})-D_5({\bf 0}))^2  + D_{55}({\bf k}) \;.
\end{eqnarray}
Thus, by adjusting the coefficients $b$, $b_1$ and $b_3$ we have
obtained  a certain elegant structure for $\Gamma_{nm}({\bf k})$ via cancellation of many terms, which by 
itself indicates that the calculations are correct and the 
transformation~(\ref{eq:W1}) -- (\ref{eq:W5}) is meaningful. Probably, we cannot do anything better. 

Nevertheless, these equations do not provide all the necessary cancellations. In particular, we can see it by evaluating
$\Gamma_{15}({\bf k})$ up to the $O \left(\epsilon^2 \right)$ order via calculation of $D_5({\bf k})$.
In this calculation we set $c=1/2$ in~(\ref{eq:Hf}), so that the Gaussian
correlation function $G_0({\bf k})$, which is related to the coupling lines in the diagram expansion, is just $k^{-2}$ within the $\epsilon$--expansion.
It is consistent with the fact that $\sum_{\bf k} {\bf k}^2  {\mid \varphi_{\bf k} \mid}^2$ is the only term,
which is included in the Gaussian
part of the Hamiltonian~(\ref{eq:Hf}), the remaining terms being treated perturbatively in this case. 
Let us denote by $\widetilde{D}_5({\bf k})$ the quantity $D_5({\bf k})$, calculated in
the usual approximation, where the fixed--point Hamiltonian is given by~(\ref{eq:Hf})
with renormalized values of the parameters. Thus, we obtain
\begin{eqnarray}
 \widetilde{D}_5({\bf k}) &=& \frac{(4!)^2 \, 5!}{2 \cdot 3!} {u^*}^2 \Dfive \label{eq:D5e} \\ 
 &=& 2 \cdot 4! \, 5! \, {u^*}^2 \Big( \sunsetz \times \sunsetk + \Dfivex \Big) + O\left( \epsilon^3 \right) \;, \nonumber
\end{eqnarray}
where $u^*$ is the renormalized value of $u$ in~(\ref{eq:Hf}), which is a quantity of order $O(\epsilon)$, and \sunsetkx is the diagram block, 
from which the constant contribution \hspace*{-1.5ex} \sunsetz is subtracted.
Here the depicted diagrams represent just the corresponding ${\bf k}$--space integrals, including
the factor $(2 \pi)^{-d}$ for each integration, all extra factors being given explicitly.
The leading singularity at $k \to 0$ in
the second line of~(\ref{eq:D5e}) is provided by first term, where $\sunsetz =\mathcal{A}$  is a positive constant and
$\sunsetk = \mathcal{A} + \frac{1}{4} K_4^2 \, k^2 \ln k$, where $K_d = S(d)/(2 \pi)^d$, $S(d)=2 \pi^{d/2} / \Gamma(d/2)$
being the surface area of $d$--dimensional sphere. This result is well known~\cite{Ma}. Hence, taking into
account that $G({\bf k}) = k^{-2} + O(\epsilon)$ holds, the considered approximation
for the renormalized Hamiltonian yields
\begin{equation}
 \Gamma_{15}({\bf k}) =  1440 \, \mathcal{A} K_4^2 {u^*}^2 \ln k  + O\left( \epsilon^3 \right)  \qquad \mbox{at} \quad k \to 0 \;.
\label{eq:ases}
\end{equation}
Thus, the desired cancellation of the terms $\propto (\ln k)^p$  in~(\ref{eq:Gamm15}) does not take place in this approximation,
at least.

In fact, the fixed--point Hamiltonian contains certain $\varphi^6$ vertex \sixx at the $O \left(\epsilon^2 \right)$ order~\cite{K2012x},
where the wave vectors of magnitude $k>\Lambda$ are related the dotted coupling line.
As a result, the diagram in the first line of~(\ref{eq:D5e}) has to be completed by one extra diagram,
where the condition $k<\Lambda$ is replaced by $k>\Lambda$ for the coupling line between two nodes
on the left hand side, to obtain the correct result for $D_5({\bf k})$ at this order of the $\epsilon$--expansion. 
Thus, the small--$k$ contribution is replaced by the large--$k$ contribution for this specific line.
As a result, this extra diagram gives no contribution to the leading singularity of $D_5({\bf k})$ at $k \to 0$,
and the asymptotic estimate~(\ref{eq:ases}) remains unaltered. Hence, the desired cancellation 
of the terms $\propto (\ln k)^p$ in $\Gamma_{15}({\bf k})$ does not take place.

\section*{Acknowledgments}
 
The authors acknowledge the use of resources provided by the
Latvian Grid Infrastructure. For more information, please
reference the Latvian Grid website (http://grid.lumii.lv).

\end{document}